\documentclass[usenatbib]{mn2e}
\usepackage{graphicx}
\usepackage{fixltx2e}
\usepackage{textcomp}

\newcommand{\Msun}{\rm{\,M_\odot}}
\newcommand{\Msunh}{\rm{\,M_\odot/h}}
\newcommand{\Mpch}{\rm{\,Mpc/h}}

\begin{document}
\title{Spin and structural halo properties at high redshift in a LCDM Universe}

\author[Davis \& Natarajan] {Andrew J. Davis,$^1$ and Priyamvada  Natarajan$^{1,2,3}$ \\
$^1$Department of Astronomy, Yale University, P.O. Box 208101, New Haven, CT 06520-8101, USA\\
$^2$Department of Physics, Yale University, P.O. Box 208120, New Haven, CT 06520-8120, USA\\
$^3$Radcliffe Institute for Advanced Study, Harvard University, 10 Garden Street, 
Cambridge, MA 02138, USA}

\maketitle
\begin{abstract}
In this paper, we examine in detail the key structural properties of
high redshift dark matter haloes as a function of their spin
parameter. We perform and analyze high resolution cosmological
simulations of the formation of structure in a LCDM Universe.  We
study the mass function, shapes, density profiles, and
rotation curves for a large sample of dark matter
haloes from $z = 15 - 6$.  We also present detailed convergence tests for individual haloes.
We find that high spin haloes have stronger clustering strengths (up to $25\%$) at all mass and redshift ranges at these early epochs.  High redshift spherical haloes are also up to $50\%$ more clustered than extremely aspherical haloes.  High spin haloes at these redshifts are also preferentially found in high density environments, and have more neighbors than their low spin counterparts.  We report a systematic offset in the peak of the circular velocity curves for high and low spin haloes of the same
mass. Therefore, estimating halo masses without knowledge of the spin, using only the circular velocity can yield errors of up to $40\%$.  The significant dependence of key structural properties on spin that we report here likely has important implications for studies of star formation and feedback from these galaxies.

\end{abstract}

\section{Introduction}
\label{sec:intro}

The currently favored model that describes the formation of structure in the Universe is the $\Lambda$ cold dark matter (LCDM) paradigm.  In this model, the initial density distribution of the Universe was nearly homogenous, with small Gaussian density perturbations imprinted during an inflationary epoch.  These fluctuations expand linearly, until the over-dense regions undergo non-linear gravitational collapse to form bound dark matter haloes.   These haloes form in a hierarchical fashion: small haloes form first, and then larger ones assemble later via merging.  In the LCDM paradigm, baryons follow the dark matter.  Since they can dissipate and cool, baryons condense, and eventually form observable galaxies in the centres of dark matter haloes.  

The properties of dark matter haloes in the context of the LCDM paradigm have been studied in detail using numerical simulations over the past couple of decades with increasing resolution \citep[e.g.][]{Davis85, Frenk88, Efstathiou88, Katz99, Kauffmann99, Bullock01B, Frenk02, Springel05}. This approach has been very fruitful in providing us with a detailed picture of the assembly and growth of structure in the Universe. These theoretical studies provide the framework within which the role of baryons and details of galaxy formation can be probed.  While collisionless dark matter in the LCDM paradigm interacts only gravitationally, baryons dissipate, have pressure, cool, form stars, and interact with radiation.  These, and other effects, introduce complications when trying to understand the properties of dark matter haloes such as their mass, angular momentum, shape, and density profiles from observations of the baryonic component.  There are, however two techniques that have allowed a more direct probe of the dark matter: gravitational lensing observations \citep[e.g.][]{Fischer00, McKay02, Hoekstra03, Mandelbaum06, Limousin07,Parker07,Evans09}, and measurements of galaxy rotation curves \citep[e.g.][]{Rubin85, Trimble87, Persic96, Salucci07}. Due to the difficulties and assumptions required to translate the observed baryonic properties to dark matter halo properties, cosmological N-body simulations offer a powerful tool to understand the properties and statistics of the dark matter haloes.

Even with dark matter only numerical simulations, much has been learned about the assembly of dark matter haloes, including the halo mass function, halo clustering, halo shape and spin at low redshift \citep[see, e.g.,][]{vitvit02, reed03, reed07, Bett07, Gao07, reed08, Maccio08, Faltenbacher09}. However, there have been few detailed studies of dark matter halo properties at high redshifts.  This is partly due to the number of particles required to resolve high redshift, low mass haloes, and still match observations of larger haloes at lower redshifts. These restrictions until recently prevented the detailed study of a statistically significant sample of collapsed haloes at high redshifts.  As the observational frontier is pushed to higher and higher redshifts with reports of the detection of galaxies out to $z \approx 7-8$ \citep{Oesch10}, a deeper understanding of the properties of the dark matter haloes that host these most distant galaxies is critical as well as extremely timely.

A few recent studies have examined specific dark matter halo properties at higher redshifts.  \citet{Heitmann06}, \citet{Warren06}, and \citet{reed07} focus on the mass function of high redshift haloes.  \citet{Moore06} trace the spatial distribution of dark matter halos from $z=12$ to the present day to understand their effect on galaxy mass haloes today.  \citet{jch01} use low resolution simulations to determine the spin and shape parameters of dark matter haloes at $z=10$.  In a recent study \citep{Davis09} we reported the results of the first high redshift and high resolution study to follow the growth of angular momentum in dark matter haloes in the mass range $10^6 \Msun$ to $10^8\Msun$ from $z=15$ to $z=6$, a period spanning 700 Myrs of cosmic time.  We found that the spin distribution at these early epochs can be fit by a log-normal distribution as at lower redshifts.  In addition, we examined the two-point correlation function of haloes and found that higher spin haloes are more clustered by factors up to $25\%$ compared to their low spin counterparts at a given mass.  This finding extended across all mass bins and redshifts in our previous study, i.e. from $10^6 - 10^8 \Msun$ and from $z=15 - 6$. 

This paper builds on our earlier work by investigating the role angular momentum and the environment play in the determination of structural properties of dark matter haloes at these epochs.  In the LCDM paradigm, haloes acquire angular momentum by tidal torques from their neighbors \citep{hoyle49, peebles69, doro70, white84}.  This picture for the acquisition and growth of angular momentum has been shown to be generally accurate in N-body simulations wherein angular momentum initially grows linearly with time \citep{barnes87} and then slows down at later times \citep{sugerman00}.  Linear theory, however, overpredicts the angular momentum when compared to the fully non-linear N-body simulations \citep{barnes87,sugerman00,porci02}.  In addition, as \citet{vitvit02} point out, linear theory predicts the angular momentum of a halo at a given redshift, but not the angular momentum of any particular progenitor at an earlier redshift.  Thus, it becomes impossible with linear theory to trace the evolution of a halo's angular momentum in a hierarchical Universe evolving via mergers.  \citet{vitvit02, maller02, Hetz06} all note that mergers do affect the spin of the halo in addition to the tidal torque model.  \citet{donghia07} study mergers and spin evolution explicitly and argue that mergers only affect the spin of unrelaxed haloes, and find that relaxed, isolated haloes show no correlation between spin and merger history.  

One way to study the acquisition of angular momentum is to correlate information about the environment with halo properties.  Previous studies have shown that halo clustering strength depends on the angular momentum of the halo at low redshifts \citep{Bett07, Gao07, Faltenbacher09}.  \citet{Avila05} find that galaxy mass haloes inside clusters have smaller spins than haloes in the field or voids, and \citet{Reed05} find low specific angular momentum in subhaloes near the central host halo.  

Observations using the Sloan Digital Sky Survey (SDSS) show that the spin parameter, 
\begin{equation}
\lambda  =  \frac{J|E_{\rm{tot}}|^{1/2}}{GM^{5/2}},
\label{eq:lambda}
\end{equation}
where $J$ is the total angular momentum, $E_{\rm{tot}}$ is the total energy, and $M$ the halo mass, has little dependence on the local environmental density \citep{cs08}.  It remains unclear, however, whether the spin parameter derived from the baryonic disk model used in interpreting SDSS data correlates well with the host dark matter halo's spin which is what is assumed.   These results are found using galaxy neighbors to trace the large scale tidal field \citep{cs09}.  As an example of the difficulties in relating baryonic properties to the host dark matter halo, \citet{Quadri03} report how slight spatial offsets between the dark matter and the baryonic disk create disturbed lensing configurations which can be easily misinterpreted if the misalignment is not included in the mass reconstruction.  Also, \citet{Mandelbaum06} report the misalignment of light ellipticity with halo ellipticity in the SDSS catalog.  These findings illustrate the  complexities when inferring dark matter halo properties from baryonic observations. 

Assembly bias refers to the observation that the clustering strength of dark matter haloes depends on an additional parameter beyond just halo mass.  Assembly bias has been studied in simulations by examining halo formation time, concentration, shape, triaxiality, velocity structure, and substructure content \citep{Harker06, Wechsler06, Bett07, Gao07, Jing07, Wetzel07, Angulo08}.  These previous works show convincingly that halo clustering depends on more than just halo mass.  However, all of these studies have been at low redshifts ($z < 5$) and for massive haloes ($M > 1 \times 10^{10} \Msun$). 

In this paper, we extend previous work by examining the formation and growth of dark matter haloes at high redshift, with an emphasis on studying the role angular momentum and environment play in regulating the structural properties of dark matter haloes.  We limit ourselves to haloes in the mass range of $10^{6} \Msun$ to a few times $10^9 \Msun$, and the redshift range $z=15$ to $z=6$.  This allows us to focus on the dark matter haloes that will likely host the first generation of stars and galaxies.  

We outline our paper as follows.  In Section \ref{sec:lowz} we summarize studies of angular momentum and assembly bias at low redshift to provide the framework for our findings at high redshift.  We describe our simulations in Section \ref{sec:sims}, and present the results of convergence tests in Section \ref{sec:CT}.  Our results from
the correlation of the spin parameter to the halo environment are
presented in Section \ref{sec:environ}, and that of the effect of
angular momentum on halo structure in Section \ref{sec:structure}.  We
conclude with a discussion of the implications of our results for high redshift galaxy formation.

\section{Previous studies at low redshift}
\label{sec:lowz}

In this section, we summarize earlier findings pertaining to measurements of halo spin and clustering to provide the context for our findings.  Many numerical simulations have shown that for massive haloes at low redshift, the distribution of the dimensionless spin parameter follows a log-normal distribution,
\begin{equation}
P(\lambda)  =  \frac{1}{\lambda \sqrt{2\pi} \sigma} \exp{\left[\frac{- (\rmn{ln}(\lambda/\lambda_0))^2}{2 \sigma^2}\right]}
\end{equation}
with typical values of $\lambda_0 \approx 0.035$ and $\sigma \approx 0.5$ \citep[e.g.,][]{donghia07, Bett07, bailin05, cole96, steinmetz95, warren92}.  

Previous studies have also shown that dark matter haloes are generally triaxial with a preference for prolateness \citep[e.g.][]{Bett07, bailin05, Faltenbacher02, cole96, warren92, Frenk88}.  \citet{Bett07} find that nearly spherical haloes have smaller spins, while there is only a weak trend of halo triaxiality with spin.  In studying halo concentration versus spin, it is seen that when unrelaxed haloes are eliminated from the sample there is only a very weak surviving (if any) correlation between these two parameters \citep{Maccio07, Neto07, Bullock01B} .

\citet{Bett07, Gao07} and \citet{Faltenbacher09} all find that haloes with larger spins are more clustered than low spin haloes at a given mass.  However, \citet{Avila05} and \citet{Reed05} find in their simulations that haloes in cluster environments have smaller spins and are more spherical than their counterparts in the field, and \citet{Hahn07} find that haloes in filaments have larger spins than haloes of  the same mass in clusters or in voids.  \citet{Maccio07} find that there is no environmental dependence on the spin parameter for haloes at a given mass.  Thus, there appears to be some question as to the extent to which the environment affects the angular momentum properties of dark matter haloes.  In this work, we explore this relationship at high redshift using both the clustering strength and the local density to characterize the environment.  

\section{Description of Simulations}
\label{sec:sims}

We run a series of N-body simulations to follow the growth of dark
matter haloes from $z\approx 100$ down to $z=6$.  We choose the particle mass such that a $10^6 \Msunh$ dark matter halo has 100 particles.  For $512^3$ particles, this requirement sets the comoving box size at $2.46 \Mpch$ and the particle mass at $M_{\rm{DM}} = 1.0 \times 10^4 \Msunh$.  The initial conditions are generated using a parallelized version of Grafic \citep{mpgrafic}, which calculates the Gaussian random field for the dark matter particles.  We use Gadget-2 \citep{Gadget05} to follow dark matter particles down to a redshift of $z=6$, with output snapshots at $z=15,12,11,10,9,8,7,6$. We use the WMAP3 \citep[\{$\Omega_{\rm{M}}, \Omega_\Lambda, \Omega_{\rm{b}}, h, n, \sigma_8$\} = \{0.238, 0.762, 0.0416, 0.732, 0.958, 0.761\},][]{wmap3} and the WMAP5 \citep[\{$\Omega_{\rm{M}}, \Omega_\Lambda, \Omega_{\rm{b}}, h, n, \sigma_8$\} = \{0.258, 0.742, 0.044, 0.719, 0.963, 0.796\},][]{wmap5} cosmological parameters for our simulations.  The WMAP3 cosmology was used for our first runs studying numerical convergence of our measurements of the angular momentum, and the WMAP5 cosmology was used for the results presented in Sections \ref{sec:environ} and \ref{sec:structure}.   Table 1 shows the runs and the relevant parameters used in this paper.  Gadget-2 uses a softening length, $\epsilon$, to soften the gravitational force to prevent spurious 2-body interactions.  

\begin{table*}
\begin{tabular}{cccccc}
\hline\hline
 Name & $N_p^{1/3}$ & L & $\epsilon$ & $M_{\rm{DM}}$ & Cosmology   \\
    & & [Mpc/h]  & [kpc/h] &  [$10^4 \Msunh$] &  \\
 \hline\hline
LoRes &256 & 2.46  & 0.3844 & 8.0 & WMAP3  \\
MedRes & 512 & 2.46 & 0.18 & 1.0 & WMAP3  \\
HiRes  & 1024 & 2.46 & 0.095 & 0.125 & WMAP3  \\
 \hline
 WM5 & 512 & 2.46  & 0.09 & 1.0 & WMAP5 \\
 \hline
 \end{tabular}
 \label{run_table}
 \caption{Catalog of runs used in this paper.  $N_p$ refers to the total number of particles, $L$ to the comoving length of the box, $\epsilon$ to the softening length, and $M_{\rm{DM}}$ to the dark matter particle mass.  The assumed cosmological parameters are ($\Omega_{\rm{M}}, \Omega_\Lambda, \Omega_{\rm{b}}, h, n, \sigma_8$) are: WMAP3 = (0.238, 0.762, 0.0416, 0.732, 0.958, 0.761); WMAP5 = (0.258, 0.742, 0.044, 0.719, 0.963, 0.796).  We note that the HiRes run achieves the $1024^3$ resolution only over one-eighth of the box, due to computational limitations. }

  \end{table*}

To identify collapsed dark matter haloes, we use the publicly available HOP code provided by \citet{HOP}.  This method groups particles with their densest neighbor.  After grouping, density thresholds are used to ensure that haloes are not being over counted due to a halo being a subhalo within a larger overdensity.  We choose the density thresholds in order to match the high redshift mass function described in \citet{reed07}.  

Once the haloes are identified, each particle in the halo is tested
to see if it is actually bound to the halo. We use SKID \citep{skid}
to do the unbinding. SKID finds the potential and kinetic energies for
all particles in a given halo and removes the most unbound particle
from the halo. Successive iterations are performed until all particles
are either bound or there are no more particles in the halo.  Without
unbinding, angular momentum properties could easily be dominated by a few transient particles not representative of the collapsed halo.  We then define the halo mass as the total mass of all the particles assigned to the halo.  Having calculated masses for our haloes, we can measure the mass function of our sample, and find that the mass function of 
haloes does not match that of \citet{reed07} at the low mass end unless particles
are unbound.  We have in our halo catalogue $\approx 24,200$ haloes in the mass range $10^{6\pm0.2}\Msun$ at $z=6$, and $\approx 16,100$ at $z=10$.  In the mass range $10^{7\pm0.2}\Msun$ we have $\approx 2,600$ haloes at $z=6$ and $\approx 1250$ at $z=10$.

\begin{figure*}[!h]
\includegraphics[scale=1,angle=0]{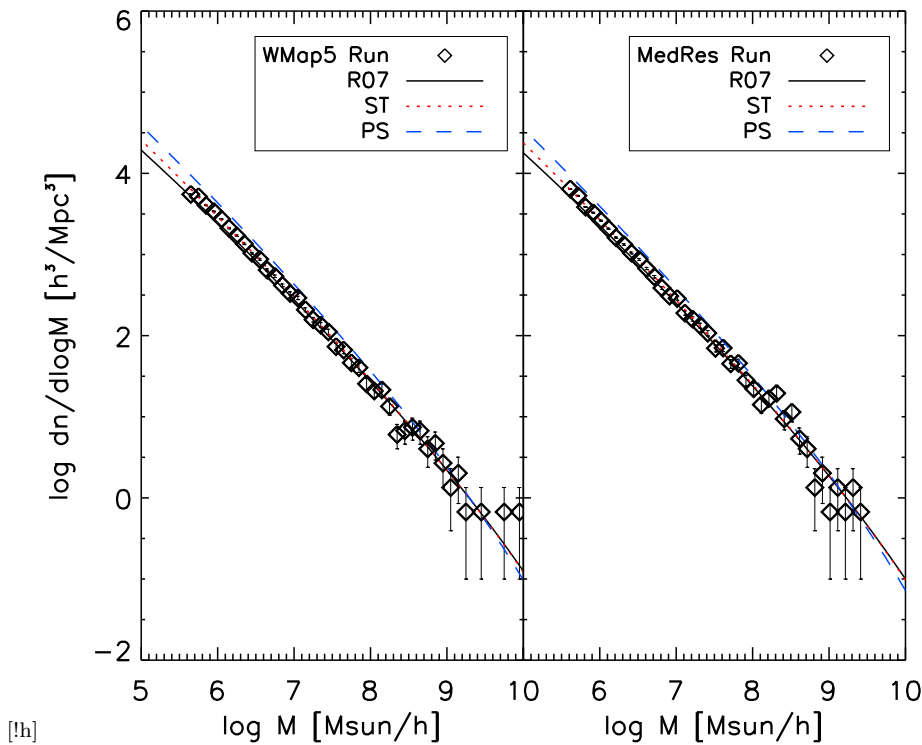}
\caption{The halo mass function for the MedRes (right) and WM5 (left) runs plotted as diamonds. The solid (black) line is the mass function from \citet{reed07}, the dotted (red) line is the Sheth-Tormen mass function, and the dashed (blue) line is the Press-Schechter mass function.}
\label{mfunc}
\end{figure*}

To ensure that our halo sample is representative, we verify the halo
mass function of our runs against theoretical predictions, as well as
against other simulations.  We use the Press-Schechter \citep{PS74} and
Sheth-Tormen \citep{st99} mass functions, and the fitting function provided by
\citet{reed07} derived from their simulations.  We show in Figure \ref{mfunc} the results from the MedRes (which uses the WMAP 3 cosmology) run and the WM5 run.  In both cosmologies, we find the mass function is poorly fit by the Press-Schechter mass function.  As the density thresholds were chosen to match the \citet{reed07} mass function, it is unsurprising that it provides a better fit than the Press-Schechter function.  However, we note that the Sheth-Tormen function provides a good fit: one that is slightly better than the \citet{reed07} function.
 
In order to calculate $\lambda$ for our halos, we must first calculate the kinetic and potential energies as well as the angular momentum from the particles.  The kinetic energy, $E_{\rm{K}}$ for a halo is the sum of the kinetic energy (\textonehalf $\:m\: \vec{v}\cdot\vec{v}$) of each particle assigned to the halo.  A halo's angular momentum, $\vec{J}$, is calculated similarly as the sum of each particle's angular momentum ($m\;\vec{r}\times\vec{v}$).  The potential energy, $E_{\rm{G}}$, of a halo is calculated using a direct summation:
\begin{equation}
E_{\rm{G}} = \sum_{i=1}^{N_h-1} \: \sum_{j>i}^{N_h} \frac{G m_i m_j}{|\vec{r}_{ij}|},
\end{equation}
where $G$ is Newton's gravitational constant, $N_h$ the number of particles in the halo, and $\vec{r}_{ij}$ is the distance between particles $i$ and $j$.  In all of these definitions, the velocities are with respect to the halo's mean velocity, and the positions are with respect to the centre of the halo.  Throughout this paper, the centre of the halo refers to the location of the densest particle, which we use as a proxy for the location of the minimum of the halo's potential well.  The only exception will be in section \ref{sec:CT}, in which we use the centre of mass to cross reference halos in simulations of different resolution.

Having now calculated $E_{\rm{K}}$ and $E_{\rm{G}}$ for our haloes, we can measure how virialized the haloes are at these early epochs.  The scalar virial theorem states that for an isolated, collisionless system in a steady state, the total kinetic energy should be equal to half the total potential energy: $2E_{\rm{K}} + E_{\rm{G}} = 0$.   Thus, measuring the total kinetic and potential energies gives us insight into the dynamical state of these dark matter haloes.  However, we note that the two key assumptions of the virial theorem (an isolated halo and steady state) are not strictly valid for these halos at these epochs.  First, these haloes are still actively merging and accreting matter, and so have an effective surface pressure that adds to the total energy of the system. In addition to the surface pressure, we also note that the haloes are not necessarily in a steady state.  The density of matter increases with redshift proportional to $(1+z)^3$, and so at a redshift of $z=6$, the mean matter density of the Universe is higher by a factor of almost $350$.  The higher density implies a larger merger rate for these dark matter haloes and hence a shorter time between mergers.  Thus haloes are less likely to be in a steady state at high redshift.  Finally, the virial theorem relates time-averaged values for the kinetic and potential energies.  However, we do not have the time resolution required for such a calculation.  We therefore use instantaneous measurements of the kinetic and potential energies, with the understanding that this will induce scatter in the energies of our haloes.

\begin{figure}
\includegraphics[scale=0.7,angle=0]{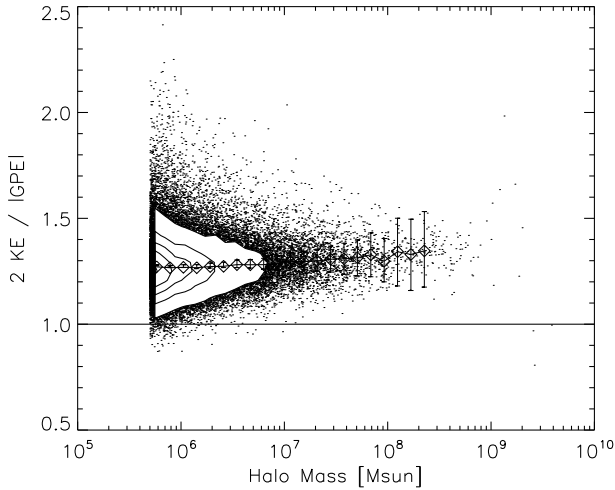}
\caption{Ratio of $2\:E_{\rm{K}}/|E_{\rm{G}}|$ for all haloes in the MedRes run at $z=6$ as a function of mass.  For fully virialized haloes, this ratio should be 1.  However, we find a mean value of 1.3, with only $0.2\%$ of haloes having a value less than unity. The diamonds depict the mean ratio as a function of halo mass.  The contours enclose ten, thirty, fifty, seventy, and ninety percent of the haloes (the same percentiles hold in all further figures with contours)}
\label{Virial}
\end{figure}

In Figure \ref{Virial}, we show the ratio of $2E_{\rm{K}}/|E_{\rm{G}}|$ for all haloes in the MedRes run at
$z=6$.  For a virialized dark matter halo, this should be approximately equal to $1$.  However, we find that the median ratio is  $1.3$, with fewer than $0.2\%$ having a virial ratio less than unity at $z=6$.  At higher redshifts, the median ratio increases slightly to $1.5$ at $z=10$.  We note that \citet{Hetz06} found similar results, albeit at lower redshifts.  These results imply that very few of our haloes are actually virialized evaluated using the above definition.   However, \citet{Bett07} labeled haloes with a ratio between $0.5$ and $1.5$ as relaxed, and \citet{Neto07} set the upper limit as $1.35$.  We find that $95\%$ of our haloes fit the \citet{Bett07} criterion, while $73\%$ fit the \citet{Neto07} criterion.  

\section{Convergence and Numerical Resolution Effects}
\label{sec:CT}

In order to verify that we have the necessary particle resolution to
capture the angular momentum properties of haloes we ran three runs that had identical initial conditions, but different resolutions.  This was achieved by creating the initial conditions for a $1024^3$ sized run (HiRes) and then lumping particles to create a $512^3$ (MedRes) and $256^3$ (LoRes) set of initial conditions.  Thus, each particle in the MedRes run has the total mass of $8$ particles in the HiRes run, and is assigned the average position and velocity of those 8 particles.  The LoRes run is similarly averaged out of the MedRes run.  We note, however, that due to computational limitations, we were unable to run the entire box at the highest resolution.  Therefore, we kept only one-eighth of the box at the highest resolution, and lumped the rest of the box to the medium resolution.  In order to avoid biases due to the interaction of differing mass particles in the HiRes run, we only kept haloes which were entirely composed of the high resolution particles.  This ensures that the haloes used for comparison are representative of the highest resolution.  We used the centre of mass to cross-match haloes across the three runs.  We match $85\%$ of the haloes between runs.  The unmatched haloes are likely cases where the group finder has joined neighboring groups with a tenuous bridge at a higher resolution that does not exist at lower resolution.  In this situation, the centre of mass would be very different between resolutions.

\begin{figure*}
\begin{center}
\includegraphics[scale=0.8,angle=0]{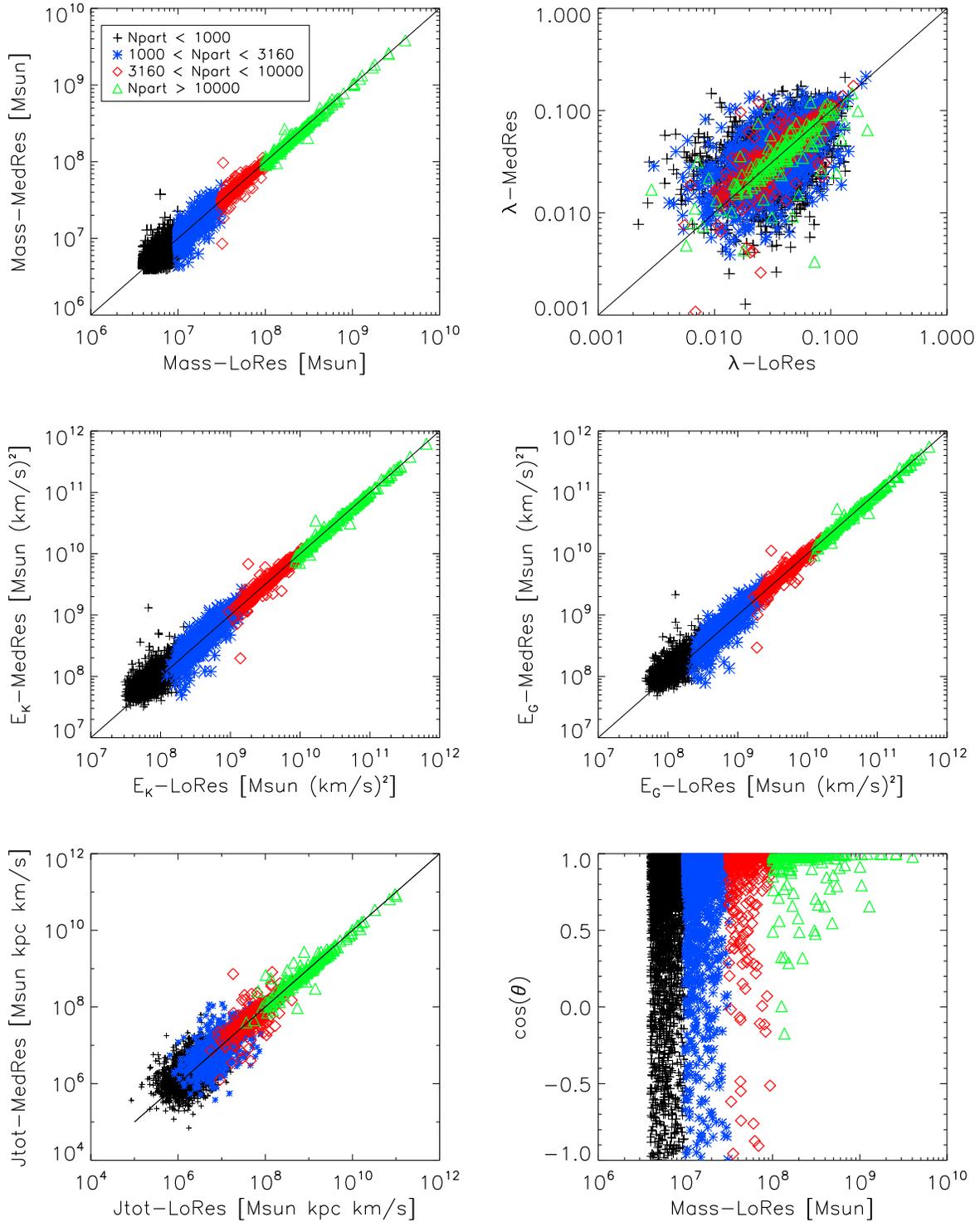}
\caption{Comparison of halo properties in the LoRes and MedRes runs.  The upper row plots haloes mass and spin, the middle row shows the kinetic and potential energies, and the bottom row shows the total angular momentum and the cosine of the angle between the angular momentum vectors in the two resolutions as a function of halo mass.  Crosses (black) refer to haloes with fewer than 1000 particles in the LoRes run, stars (blue) to haloes with fewer than 3160 particles, diamonds (red) to haloes with fewer than 10000 particles, and triangles (green) to haloes with more than 10000 particles.}
\label{CT_test}
\end{center}
\end{figure*}

In Figure \ref{CT_test}, we compare the halo mass, spin, kinetic energy, potential energy, and total angular momentum between the LoRes and MedRes simulations.  The bottom right panel shows the cosine of the angle between the angular momentum vectors in the two simulations.  Table \ref{Comp_stats} lists the mean and standard deviation of the fractional difference of those same quantities as a function of number of particles in the LoRes run.  We also include the offset in the centre of mass position, which was used to cross-match the haloes.  We find good agreement in the masses of individual haloes, with a mean fractional difference of $14\%$ at the lowest particle resolution (haloes with less than $1,000$ particles).  However, there is some spread in calculating the kinetic and potential energies, with differences up to $30\%$ in the smallest bin.  The angular momentum has the largest variation of the quantities used to calculate $\lambda$.  For the smallest haloes, we find a mean fractional difference of $73\%$, which only decreases to $25\%$ in the largest haloes (haloes with more than $10,000$ particles.   We also find that the direction of the angular momentum vector is biased at low particle resolution.  

These findings lead to a large spread in the spin parameter for the same haloes at differing resolution.  There does not appear to be any systematic offset in $\lambda$ between resolutions, as the top right panel of Figure \ref{CT_test} shows.  This is unlike \citet{Trenti10} who found that for haloes with less than 100 particles, the spin parameter measurement is biased high at lower resolution.  While we do not show the comparison between the MedRes and the HiRes subregion, we find the same trends with increasing resolution.  One possible explanation for the larger dispersion in $\lambda$ may be the difficulty in defining an outer edge or boundary for dark matter haloes.  A particle at the edge adds little to the halo's mass, but it will add a considerable amount of angular momentum.  Thus, a measurement of any particular halo's angular momentum or spin parameter will be dependent on the halo finding algorithm used in the study \citep{Trenti10}.

\begin{figure}
\centering
\includegraphics[width=0.5\textwidth]{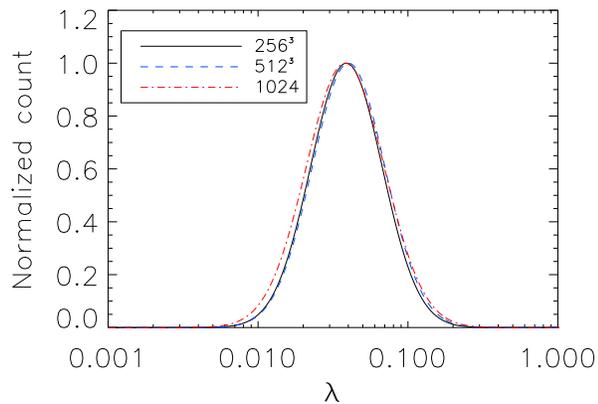}
\caption{Comparison of the spin distribution of all three resolution runs.  The solid (black) curve shows the log-normal fit for the MedRes run, the dashed (blue) curve is the fit for the LoRes run, and the dot-dashed (red) curve is the fit for the HiRes run.  Note that while the spin values for individual haloes have not converged, the statistical ensemble shows little variance between the three different resolutions.}
\label{CT_ensemble}
\end{figure}

In addition to the boundary issue, particle discreteness and resolution is also problematic when measuring halo properties such as spin and angular momentum.  As seen in Table \ref{Comp_stats}, the dispersion decreases with increasing number of particles across all quantities.  This implies that a large source of the scatter is due to particle resolution.   

The ensemble properties of the haloes' spin parameter reaches convergence at the medium resolution.  Figure \ref{CT_ensemble} shows the log-normal fit to the spin distribution for the three runs. The three runs have a mean $\lambda$ of $0.039$ (LoRes), $0.040$ (MedRes), and $0.038$ (HiRes).  In the further analysis reported below, we use the results from the WM5 run, which has the same mass resolution as the MedRes run.    

\begin{table*}
\caption{Statistical comparison of our LoRes and MedRes runs.  We present the mean and standard deviation of seven quantities: the offset between the centre of mass (in physical kpc); the fractional difference (defined as $|X_m-X_l|/X_l$ for property $X$) in mass, spin, kinetic energy, potential energy, and total angular momentum; and the cosine of the angle between the angular momentum vectors in the two runs.  We find that the scatter decreases with increasing resolution for all halo properties, and that the angular momentum vector is more robust at large particle numbers.}
\begin{tabular}{|c|rr|rr|rr|rr|}
\hline
\hline
 Property & \multicolumn{2}{|c}{$N < 1000$} & \multicolumn{2}{|c}{$1000 < N < 3160$} & \multicolumn{2}{|c}{$3160< N < 10000$} & \multicolumn{2}{|c|}{$N > 10000$} \\
 & Mean & $\sigma$ & Mean & $\sigma$ & Mean & $\sigma$ & Mean & $\sigma$ \\ \hline
 $\Delta r$ & 0.207 & 0.162 & 0.185 & 0.172 & 0.158 & 0.173 & 0.171 & 0.198 \\
 $\Delta M/M_l$ & 0.141 & 0.176 & 0.107 & 0.114 & 0.072 & 0.119  & 0.046 & 0.066 \\ 
 $\Delta \lambda/\lambda_l$ & 0.563 & 0.873 & 0.422 & 0.692 & 0.257 & 0.409 & 0.254 & 0.532 \\ 
 $\Delta E_{\rm{K}}/(E_{\rm{K}})_l$ & 0.309 & 0.483 & 0.168 & 0.166 & 0.102 & 0.163 & 0.097 & 0.226 \\ 
 $\Delta E_{\rm{G}}/(E_{\rm{G}})_l$ & 0.286 & 0.426 & 0.157 & 0.160 & 0.094 & 0.158 & 0.102 & 0.230 \\ 
 $\Delta J_{\rm{tot}}/(J_{\rm{tot}})_l$ & 0.734 & 1.39 & 0.541 & 1.24 & 0.377 & 1.85 & 0.248 & 0.584 \\ 
 $cos(\theta)$ & 0.598 & 0.475 & 0.745 & 0.388 & 0.857 & 0.300 & 0.916 & 0.179 \\ \hline
\end{tabular}
\label{Comp_stats}
\end{table*}

\section{Spin and Environment}
\label{sec:environ}

\begin{figure*}
\includegraphics[scale=0.6,angle=0]{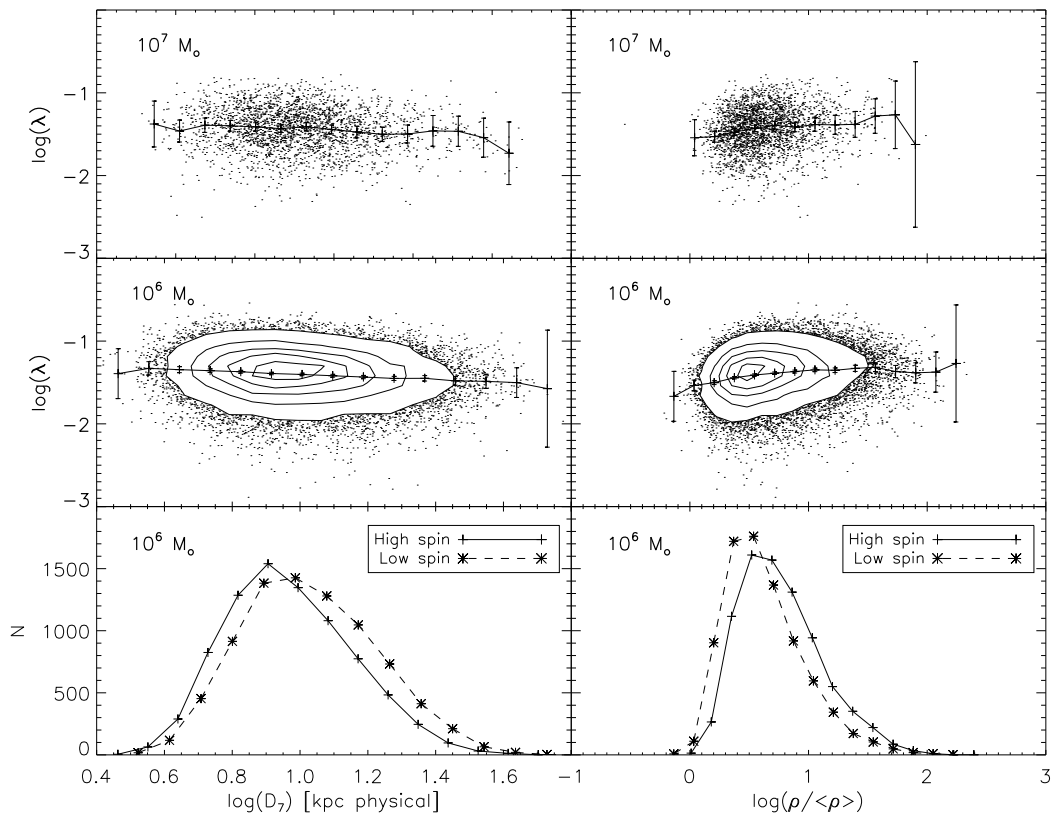}
\caption{Two measures of environment versus spin parameter:  the distance to the $7^{th}$ nearest neighbor ($D_7$, left column), and the over-density within $10 R_{178}$ (right column).  The solid curves show the mean spin, and the error bars reflect Poisson noise. The top two rows show two different mass bins.  The bottom row plots histograms of the $10^6\Msun$ halo sample, binned by spin such that one third of the haloes lie in each of the high and low spin bins.  We find only weak trends in each mass bin, but they do confirm our main finding from \citet{Davis09} that higher spin haloes are more likely found in denser environments as they have smaller values of $D_7$ and reside in over-dense regions.}
\label{environ}
\end{figure*}

In our previous work \citep{Davis09}, we found that at high redshift, higher spin haloes are more clustered than lower spin haloes at a given mass, using the halo-halo correlation function as a proxy for the environment.  As in our previous work, we again define distinct spin bins from the lognormal distribution, each with roughly one third of the haloes, and denote them as high, median, and low spin bins.  Using several different ways to characterize the environment, we report our findings of correlations between measures of the environment with the spin parameter for haloes in the mass range $10^6 \Msun$ to $10^8 \Msun$.  

We first use the distance to close neighbors as a measure of environment.  This distance is calculated as the separation between each halo's centre (i.e. the densest particle).  We measure both the distance to the $3^{rd}$ ($D_3$) and $7^{th}$ ($D_7$) nearest neighbors.  Both distances show the same result: haloes with closer neighbors have a small tendency to have larger spins.  The second method that we use to measure environment is the over-density within ten times the virial radius.  We use as a proxy for the virial radius $R_{178}$: the radius of a spherical region which has an average density of $178 \rho_{crit}$.   The factor $178$ is the overdensity criterion $\Delta(z)$ evaluated at $z=6$.  \citet{Bryan98} give a fitting formula in a LCDM universe for $\Delta(z)$, but at high redshifts, $\Delta(z)$ is approximately the same as it would be in an Einstein-de Sitter universe, such that $\Delta(z) = 18\pi^2 \approx 178$. To calculate this radius, we use all particles surrounding the centre of the potential for the given halo, regardless of whether or not the particles are part of the halo.  This is to account for the fact that not all haloes extended out far enough to reach the required density threshold.  In other words, some haloes have $R_{178}$ outside the most distant particle assigned to the group.  We then calculate the over-density within $10 R_{178}$, and find that haloes in over-dense regions have a slight tendency to have higher spins.  Figure \ref{environ} shows the trends with spin for both measures of environment in two mass bins: $10^{6\pm0.2} \Msun$ and  $10^{7\pm0.2} \Msun$, and histograms for the lowest mass bin after separating our sample by spin.  The histograms for the $10^7 \Msun$ and $10^8 \Msun$ bins show qualitatively the same trends. 

We find no strong correlation between environment and spin -- it is possible to find low spin haloes in very dense environments, and high spin haloes in sparse environments.  However, we do find a small excess probability that haloes with lower spins are in less dense environments, as evidenced by the histograms in Figure \ref{environ}.   This is in contrast to \citet{Avila05}, who report that galaxy scale haloes (masses $< 5 \times 10^{11} \Msunh$) in clusters have lower spin than isolated haloes in the field.  Interestingly, they find that the trend they report also disappears as they go back to $z=1$.  This may imply a trend with redshift where at high redshift, high spin haloes are in denser environments, and at very low redshift, high spin haloes are in less dense environments.  However, a strict comparison of these two numerical studies cannot be made  as we have significantly more haloes, probe significantly higher redshift slices and significantly lower halo masses.  \citet{Avila05} only use the over-density criterion to quantify the environment, whereas our findings hold across three different measures of environment.  We note that the trends reported here have shallow slopes, similar to the \citet{Maccio07} work.  We conclude that the dependence of spin on environment is weak for the halo masses and epochs studied here.

\section{Structural properties}
\label{sec:structure}
\subsection{Concentration}
\label{ssec:concentration}

In determining the effect of halo spin on halo structural properties, we first turn to the mass distribution within the halo.  We fit the NFW profile,
\begin{equation}
\rho(r)_{\rm{NFW}} = \frac{\rho_0}{r/r_{\rm{s}}\;(1+r/r_{\rm{s}})^2}
\end{equation}
\citep{NFW96, NFW97}, to each halo through a least squares fit to the radially averaged density profile calculated using all particles assigned to the halo by our group finding algorithm.   In doing the least squares fit, we seek to minimize $\chi^2$, which is defined as
\begin{equation}
\chi^2 = \frac{1}{\nu}\sum_{i}^{N_{\rm{bins}}}[\log \rho - \log \rho_{\rm{NFW}}]
\end{equation}
and where $\nu$ is the number of degrees of freedom.  We then use the concentration parameter, defined as $C_{178} = R_{178}/r_{\rm{s}}$, to characterize the density profile.  In Figure \ref{Struct} (left column) we show the measured concentrations versus spin parameters binned by mass.  There is only a small trend with spin, implying that the density structure has only a weak dependence on the angular momentum properties of the halo.  In the bottom left panel of Figure \ref{Struct}, we show the histogram of concentration values for haloes in the $10^6 \Msun$ mass bin.  We report little to no dependence of $C_{178}$ on spin in these mass ranges.   We also look at the relationship between halo mass and $C_{178}$.  We find that the fitting function in \citet{Bullock01A} extrapolated down to these masses and up to these high redshifts, given by $C_{\rm{vir}} \approx 9 \mu^{-0.13}/(1+z)$, where $\mu = M/M_\star$ and $M_\star$ is the typical collapsing mass at redshift $z$, is not a good fit, and predicts values for the concentration that are too small compared to our estimates from the simulation.  This disagreement is not surprising as Bullock et al. studied higher mass haloes at  lower redshifts ($z\,<\,5)$ that have had a significant amount of time to merge and virialize.  

\subsection{Circular Velocity}
\label{ssec:vcirc}

\begin{figure*}
\includegraphics[scale=0.6]{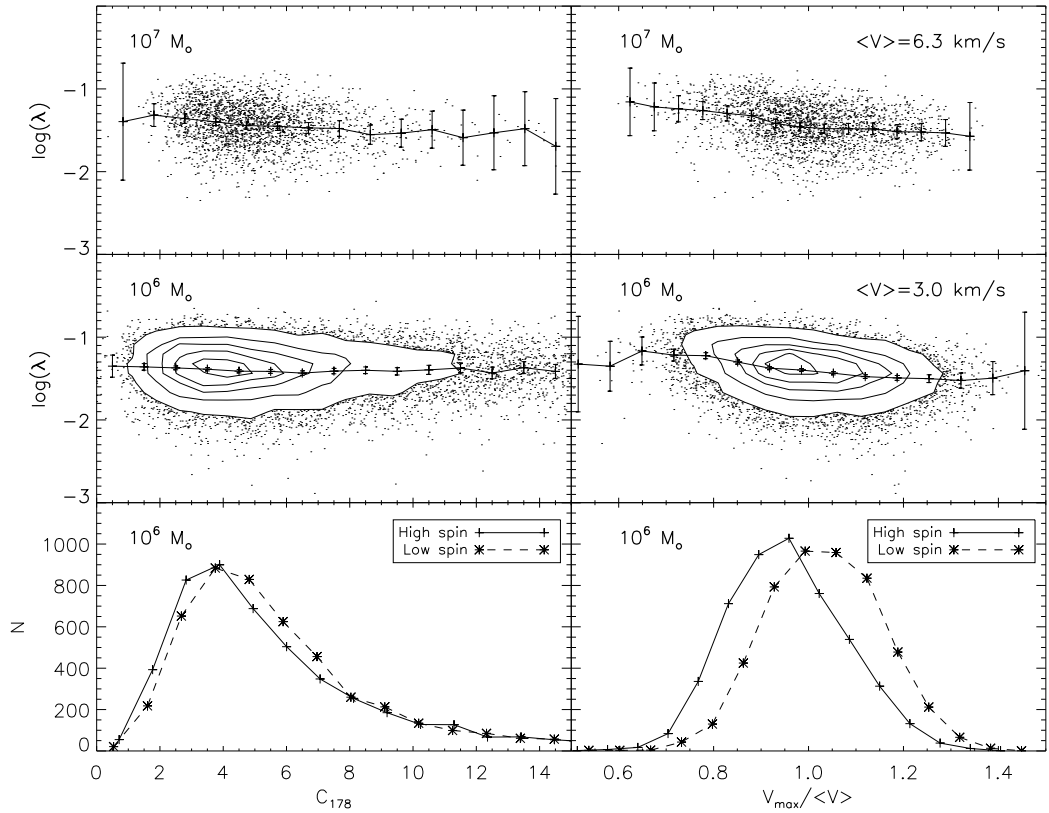}
\caption{Concentration (left column) and $V_{\rm{max}}/<V>$ (right column) versus spin for haloes binned by mass.  $V_{\rm{max}}$ is the maximum value of the circular velocity curve, given by $V_c=\sqrt{GM(<r)/r}$, and $<V>$ is the mean value of $V_{\rm{max}}$ for all haloes in the given mass bin. 
The bottom row shows histograms of the $10^6 \Msun$ haloes, binned again by spin.  While the concentration does not vary much with spin, the peak of the rotational curve does vary with spin.  This implies that deriving masses from observed velocity curves will have an additional  systematic error due to the spin of the host dark matter halo.  There is a difference of 20\% in the peak value for $10^6 \Msun$ haloes and a 10\% spread for the $10^7 \Msun$ haloes with spin parameter.}
\label{Struct}
\end{figure*}

Observed rotation curves are used to obtain mass estimates of galaxies and clusters.  The measured velocity of stars and gas reflects the combined gravitational potential of both the baryons and dark matter in the galaxy.  To infer the total mass from an observationally determined rotation curve requires understanding how both types of matter are distributed spatially in galaxies.  \citet{Persic96} report that for Sb-Im spirals, rotation curves can be represented by a universal function which is the sum in quadrature of two velocity curves: one from the disc and one from the dark matter halo.  This model is developed further in \citet{Salucci07},  where the dark matter velocity component is strictly a function of the virial mass of the halo.  If, however, the velocity curve of the dark matter halo is dependent on a second parameter, such as $\lambda$, we would expect to find a systematic error in a mass estimate for the galaxy derived from the observed rotation curve.  

A relationship between $\lambda$ and a characteristic circular velocity can be expected if we consider an alternative spin parameter given in \citet{Bullock01B}: $\lambda^\prime = J/\sqrt{2}MVR$, where $J$ is the angular momentum inside some radius, $R$, $M$ is the mass inside $R$ and $V$ is the circular velocity at $R$.  For the case of a truncated, singular isothermal sphere, $\lambda^\prime = \lambda$ at the virial radius.  Thus it is expected that a relationship between the $\lambda$ used in our work and a characteristic circular velocity should exist.

We find that the peak velocity, $V_{\rm{max}}$, in the measured circular velocity curve, defined as $V_c = \sqrt{GM(<r)/r}$, systematically depends on spin.  To measure this curve, we sort all particles in the halo by their distance to the halo centre.  From this, we can read off the circular velocity curve at the radius of each particle, and we define $V_{\rm{max}}$ as the maximum value of this curve.  Figure \ref{Struct} (right column) shows that for all mass bins, there is a systematic offset in the peak velocity as a function of halo spin: high spin haloes have higher $V_{\rm{max}}$ than their low spin counterparts.  The bottom panel shows two histograms of $V_{\rm{max}}$, binned by spin.  The peak of the high and low spin curves is offset by 20\% in the $10^6 \Msun$ mass bin, and 10\% in the $10^7 \Msun$ haloes (not plotted).  

From a given $V_{\rm{max}}$ one can infer a halo mass using the definition of circular velocity such that $M(<r) = V_{\rm{max}}^2  r / G$.  Alternatively, one can use published relationships between $V_{\rm{max}}$ and $M_{\rm{vir}}$ such as the power law $V_{\rm{max}} \propto M^{1/\alpha}$ where $\alpha = 0.31 \pm 0.08$ from \citet{Shaw06}  \citep[see also][]{Kravtsov04, Hayashi03,  Bullock01A}.  Either way, when two halos of the same (unknown) mass have different spins, there is a bias in the resulting $V_{\rm{max}}$, which will lead to a bias in the mass estimated from the circular velocity.  \textbf{Therefore two observed galaxies with differing values of $V_{max}$ could in fact inhabit dark matter haloes of the same mass but with different spin parameters.  Alternately, two galaxies with the same value of $V_{\rm{max}}$ could have different masses due to their different spin values.}   We can estimate the extra uncertainty in the halo mass due to the bias induced by $\lambda$.  Using the circular velocity definition, we can translate a difference in velocity to a difference in mass:  $2 \Delta(v) / v = \Delta(M) / M$.  Therefore the estimated total mass from an observationally measured value of $V_{\rm{max}}$ can be off by 20-40\% arising due to the unknown value of the spin parameter of the dark matter halo that hosts the galaxy.

\subsection{Halo Shape and Spin}

Another well explored correlation at low redshift is between the shapes of dark matter haloes and the spin parameter. Here, we investigate the corresponding correlations for these high redshift haloes. The existence of a correlation between halo shape and spin is likely to have important consequences for the formation of pre-galactic disks in these haloes.

\begin{figure*}
\includegraphics[scale=0.6]{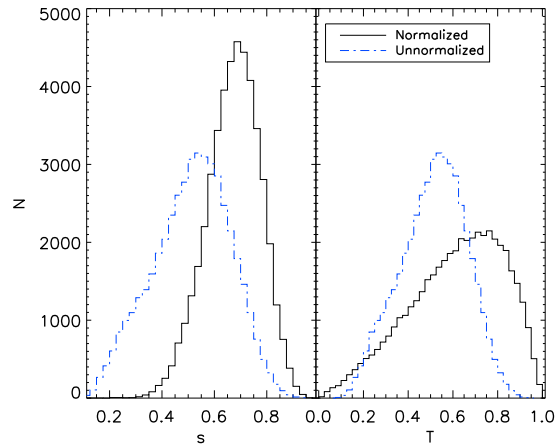}
\caption{Histograms of sphericity, $s$ (left), and triaxiality, $T$ (right), for all haloes in the WM5 run.  We show the values of $s$ and $T$ found using a normalized (solid black curve) and an unnormalized (dashed blue curve) moment of inertia tensor. We find that using the unnormalized tensor gives significantly more aspherical haloes and fewer prolate haloes than using the normalized tensor.}
\label{NormShape}
\end{figure*}

We calculate the sphericity, $s = c/a$, and triaxiality, $T=(a^2-b^2)/(a^2-c^2),$ from the eigenvalues ($ a > b > c$) of the normalized moment of inertia tensor, \[I_{ij} = \sum_{n}{\frac{x_i x_j}{|\vec{x}|}},\] where $\vec{x}$ is the distance to the halo center.   The tensor is calculated using only the particles assigned to the halo, not from all particles within the $R_{178}$ as is often done.  This allows for consistency with our measurements of $\lambda$.  We used the normalized tensor so that we do not weight particles on the outskirts of a halo stronger.  This helps to prevent the halo shape being dominated by residual tidal features from a recent merger or other irregularities found only in the outer regions which are not present in the halo interior.  We caution that both the normalized \citep[e.g.][]{Avila05, Allgood06} and unnormalized \citep[e.g.][]{jch01,Shaw06, Maccio08, Faltenbacher09} tensor are used in the literature, and so caution must be used when comparing shape distributions.  We show in Figure \ref{NormShape} histograms of $s$ and $T$ when calculated using both a normalized and unnormalized moment of inertia tensor.  We find that many more halos are significantly aspherical, and halos tend to be less prolate when the shape is found using the unnormalized tensor.  This is contrary to the study in \citep{Allgood06}, which reported no systematic offset in $s$ between the two methods.  We note, however, that they only used particles within $0.3 R_{\rm{vir}}$, rather than all particles assigned to the halo as we do.  Therefore, our unnormalized method will have a stronger bias because we include distant particles in the calculation of the halo shape.  In what follows, we use the normalized tensor to calculate the shape of the haloes.  

\begin{figure*}
\includegraphics[scale=0.6]{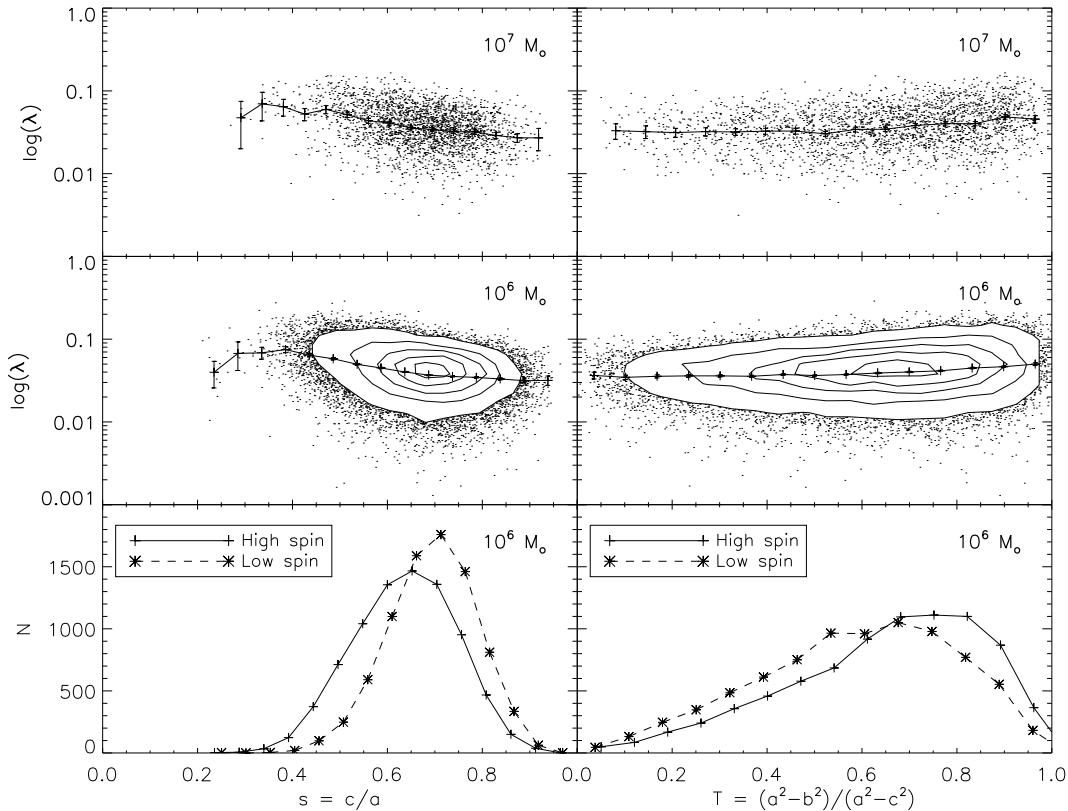}
\caption{Sphericity, $s$ (left panels), and triaxiality, $T$ (right panels), versus spin parameter for haloes binned by mass.  High spin haloes are less circular than their low sin counterparts.  High spin haloes are also more likely to be prolate.}
\label{shape}
\end{figure*}

\begin{figure*}
\includegraphics[scale=0.6,angle=90]{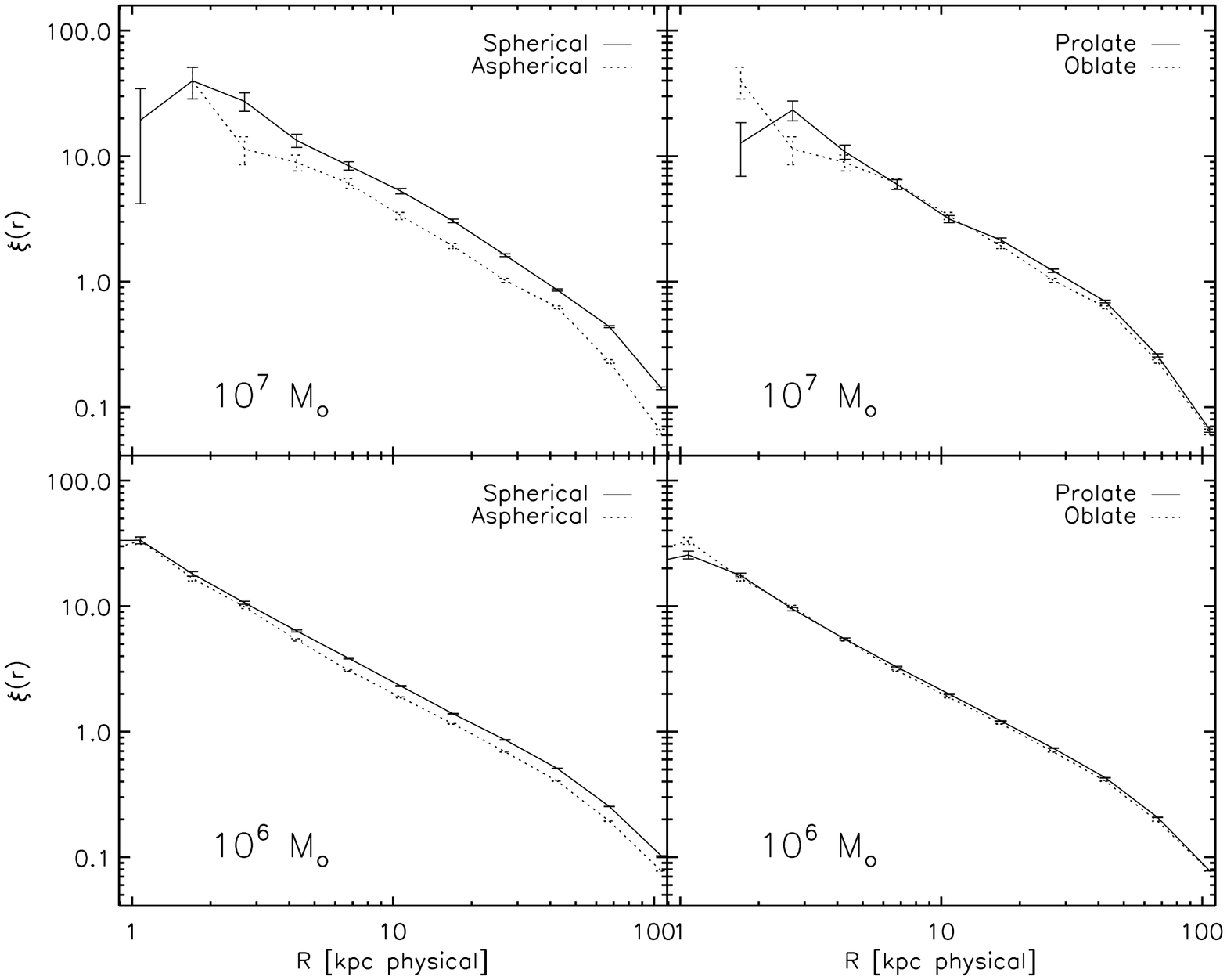}
\caption{Correlation function for haloes binned by their sphericity (left), triaxiality (right) and mass.  We find that nearly spherical haloes are more clustered than aspherical haloes, by $50\%$ in the $10^7\Msun$ bin and by $20\%$ in the $10^6\Msun$ bin.  We find no difference in clustering strength when separating haloes based on $T$.}
\label{corr_shape}
\end{figure*}

We show in Figure \ref{shape} the relations between $s$ (left column), $T$ (right column), and $\lambda$, as well as the histograms of $s$ and $T$ binned by spin and mass.  For higher spin haloes, the trend is for haloes to become less spherical, and more prolate.  This is in agreement with studies at low redshift and higher masses \citep{Maccio08,Bett07,Allgood06,Avila05}.

We also plot the two point correlation function, $\xi(r)$ of the haloes in Figure \ref{corr_shape}, binning the haloes by their values of $s$ and $T$.  The correlation function represents the excess probability of finding a halo at a distance $r$ when compared to a random distribution of haloes.  We use the same method as our previous work \citep{Davis09} to calculate $\xi(r)$.  We calculate $\xi(r)$ for haloes in four different bins, corresponding to large and small values of $s$ (left column of Figure \ref{corr_shape}) and $T$ (right column) .  The cuts were chosen so that one third of the haloes lie in each bin.  We also separate out haloes by mass so that we eliminate any mass effect on the value of $\xi(r)$.  In the top row of Figure \ref{corr_shape} we show $\xi(r)$ for haloes with $M = 10^7 \Msun$ and in the bottom row we show $10^6 \Msun$ haloes.

We find that nearly spherical haloes (values of $s$ close to 1) are more clustered than the more aspherical haloes.  This trend is stronger for haloes in the $10^7 \Msun$ mass range than in the $10^6 \Msun$ mass range.   In the higher mass bin, we find an increase of $50\%$ in the correlation of nearly spherical haloes when compared to the extremely aspherical halo sample.  In the lower mass range, we report a smaller increase of only $20\%$.   Thus we conclude that our results at high redshift follow trends found at lower redshift \citep{Faltenbacher09,Bett07,Avila05}.

Unlike the sphericity parameter, when we separate haloes into high and low values of $T$, we see little change in $\xi(r)$.   This is somewhat surprising, as \citet{Faltenbacher09} find at low redshift an offset in $\xi(r)$ when binning haloes by $T$. Finally, we note that the clustering strength is weaker if an unnormalized moment of inertia tensor is used to calculate $s$ and $T$ of the haloes.  

\begin{figure*}
\includegraphics[scale=0.5, angle=90]{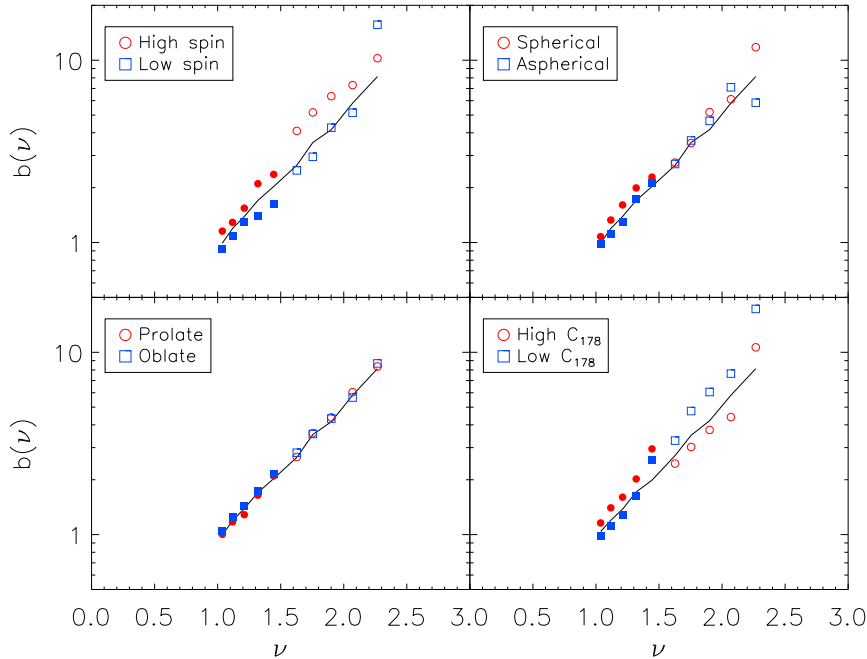}
\caption{Bias as a function of peak overdensity, $\nu(M,z)$.  We calculate the bias at two redshifts: $z=6$ (filled symbols) and $z=10$ (open symbols).  The top left panel bins haloes by their spin parameter, the top right by their sphericity, $s$, the bottom left by the triaxiality, $T$, and the bottom right by the concentration, $C_{178}$.  We see a strong offset in bias due to spin and concentration, a weaker one due to $s$, and no offset due to $T$.}
\label{bias}
\end{figure*}

In Figure \ref{bias}, we show the bias parameter, $b = \sqrt{\xi_{\rm{MM}}/\xi_{\rm{HH}}},$ as a function of the halo peak height $\nu(M,z) = \delta_c / (\sigma(M) D(z))$, where $D(z)$ is the growth function \citep{mo02}.  This allows us to compare results from two redshifts ($z=10$ and $z=6$) and from 5 mass bins at each redshift.  We split our haloes according to four properties: $\lambda$, $s$, $T$, and $C_{178}$.  We report that the spin parameter has the largest effect on the bias, while the triaxiality has the least.  This implies that angular momentum has the strongest dependence on environment of all these considered variables and will be the source of the largest systematics due to differing environments between haloes.  Therefore, baryonic properties that depend on the angular momentum of the host halo should have systematic offsets due to their local environment.  This may play a large role in the formation of the earliest galactic disks, as well as affect semi-analytic models which relate the baryonic spin to the dark matter halo spin \citep[][e.g.]{Croton06,Benson10}, and estimates of the dark matter halo spin from the observed baryonic disk \citep{cs08,cs09}.

\section{Discussion and conclusions}
\label{sec:disc}

Our key findings can be summarized as follows:
\begin{itemize}
\item We have measured the spin, concentration, circular velocity, sphericity, and triaxiality parameter for a statistically large sample of dark matter haloes at high redshift ($z > 6$).  
\item High spin haloes at high redshift are $25\%$ more clustered than their low spin counterparts at a given mass, and are more likely to be found in high density environments.
\item High spin haloes (with masses $\le 10^7\Msun$) have smaller maximum circular velocities than low spin haloes, leading to errors up to $40\%$ in the derived enclosed mass.
\item High spin haloes at high redshift are more likely to be aspherical and prolate, similar to findings at low redshift.
\item Nearly spherical haloes are up to $50\%$ more clustered than extremely aspherical haloes, while there appears to be no difference in the clustering strength based on the triaxiality of the haloes.  
\end{itemize}

Our findings have an impact in two general areas: the role of angular momentum in halo structure and formation, and the role of assembly bias at high redshift.  Our findings show that angular momentum has a measurable correlation with structural properties, including the concentration, sphericity, and triaxiality.  Also, haloes with higher spin are preferentially found in higher density environments.  The finding that halo spin correlates with local environment at high redshift is important  to the understanding of the evolving properties of the baryonic component of dark matter haloes. A correlation between spin and baryonic properties, such as formation time, disk rotational speed, or disk size, would be the specific consequences of the 
 correlation with the environment.  These correlations are likely to be significantly stronger at high redshift, before too many mergers have happened which could destroy any correspondence between the dark matter spin and the baryonic structure.  

The role of assembly bias in halo evolution has been discussed at low redshifts.  Our work extends the study of assembly bias to high redshifts, when the first galaxies form.  We see similar results to those at low redshift when studying the dependence of clustering on spin and sphericity of dark matter haloes.  However, our results differ when looking at the triaxiality of haloes.  We find that there is no difference in clustering strength between prolate and oblate haloes.  

Our findings are of particular importance now that galaxies are being found at these high redshifts \citep{Oesch10}.  The clustering of galaxies has been used to infer the masses of their host dark matter haloes \citep[e.g., ][]{Quadri08}.  However, we find that properties other than mass - in particular halo spin - affect the measured correlation function.  This additional parameter will induce errors in mass estimates for  dark matter haloes inferred purely from clustering measurements.  This fits in with the results of \citet{Quadri08}, who suggest that mass is likely not the only parameter that drives the interaction of haloes with their large-scale environment.  

In addition to mass measurements, assembly bias will play an important role in feedback at these high redshifts.  At the highest redshifts, simulations show that Population III stars have a large impact on their environment due to radiative and supernova feedback \citep{Johnson07, grief07, Whalen08A, Whalen08B}.  One consequence of our findings is that if angular momentum affects the formation and evolution of these Pop III stars, their feedback effects will show an environmental bias.  Thus, the distribution of metals and reionization will be more clustered than otherwise expected.  We also expect that because these stars are the first baryonic objects to collapse, their properties can be expected to have a stronger relationship to their host dark matter halo than galaxies today, which have undergone multiple mergers.  We intend to pursue the consequences of our findings on baryonic results in future work.  Our results suggest that the angular momentum properties of dark matter haloes likely have consequences for the properties of the first stars and galaxies hosted by them.

\section*{Acknowledgments}
This work was supported in part by the facilities and staff of the Yale University Faculty of Arts and Sciences High Performance Computing Center.  We also wish to thank Zheng Zheng, Laurie Shaw, Jason Tumlinson, Elena D'Onghia, and Lars Hernquist for helpful discussions and comments.


\begin{thebibliography}{}

\bibitem[\protect\citeauthoryear{Allgood et~al.}{2006}]{Allgood06}
Allgood, B., Flores, R., Primack, J., Kravtsov, A., Wechsler, R., Faltenbacher, A., \& Bullock, J. 2006, MNRAS, 367, 1781

\bibitem[\protect\citeauthoryear{Angulo, Baugh, \& Lacey}{2008}]{Angulo08}
Angulo, R., Baugh, C., \& Lacey, C. 2008, MNRAS, 387, 921

\bibitem[\protect\citeauthoryear{Avila-Reese et~al.}{2005}]{Avila05}
Avila-Reese, V., Colin, P., Gottlober, S., Firmani, C., \& Maulbetsch, C. 2005, ApJ, 634, 51

\bibitem[\protect\citeauthoryear{{Bailin} \& {Steinmetz}}{2005}]{bailin05}
{Bailin}, J., \& {Steinmetz}, M. 2005, ApJ, 627, 647

\bibitem[\protect\citeauthoryear{{Barnes} \& {Efstathiou}}{1987}]{barnes87}
{Barnes}, J. \& {Efstathiou}, G. 1987, ApJ, 319, 575

\bibitem[\protect\citeauthoryear{{Bett} {et~al.}}{2007}]{Bett07}
Bett, P., Eke, V., Frenk, C., Jenkins, A., Helly, J., Navarro, J. 2007, MNRAS, 376, 215

\bibitem[\protect\citeauthoryear{Benson \& Bower}{2010}]{Benson10}
Benson, A., \& Bower, R. 2010, MNRAS accepted, arXiv:1003.0011v1

\bibitem[\protect\citeauthoryear{Bryan \& Norman}{1998}]{Bryan98}
Bryan, G., \& Norman, M. 1998, ApJ, 495, 80

\bibitem[\protect\citeauthoryear{Bullock et~al.}{2001A}]{Bullock01A}
Bullock, J., et~al. 2001, MNRAS, 321, 559

\bibitem[\protect\citeauthoryear{Bullock et~al.}{2001B}]{Bullock01B}
Bullock, J., Dekel, A., Kolatt, T., Kravtsov, A., Klypin, A., Porciani, C., \& Primack, J. 2001, ApJ, 555, 240

\bibitem[\protect\citeauthoryear{{Cervantes-Sodi} {et~al.}}{2008}]{cs08}
{Cervantes-Sodi}, B., {Hernandez}, X., {Park}, C., \& {Kim}, J., 2008 MNRAS, 388, 863

\bibitem[\protect\citeauthoryear{Cervantes-Sodi, Hernandez \& Park}{2010}]{cs09}
Cervantes-Sodi, B., Hernandez, X., \& Park, C. 2010, MNRAS, 402,1807

\bibitem[\protect\citeauthoryear{{Cole} \& {Lacey}}{1996}]{cole96}
{Cole}, S., {Lacey}, C. 1996, MNRAS, 281, 716

\bibitem[\protect\citeauthoryear{Croton et~al.}{2006}]{Croton06}
Croton, D., et~al. 2006, MNRAS, 365, 11

\bibitem[\protect\citeauthoryear{Davis et~al.}{1985}]{Davis85}
Davis, M., Efstathiou, G., Frenk, C., \& White, S. 1985, ApJ, 292, 371

\bibitem[\protect\citeauthoryear{Davis \& Natarajan}{2009}]{Davis09}
{Davis}, A., \& {Natarajan}, P. 2009, MNRAS, 393, 1498

\bibitem[\protect\citeauthoryear{{D'Onghia} \& {Navarro}}{2007}]{donghia07}
{D'Onghia}, E., \& {Navarro}, J. 2007, MNRAS, 380, 58

\bibitem[\protect\citeauthoryear{{Doroshkevich}}{1970}]{doro70}
{Doroshkevich}, A.~G. 1970, Astrophysics, 6, 320

\bibitem[\protect\citeauthoryear{Dunkley et~al.}{2009}]{wmap5}
Dunkley, J., et~al. 2009, ApJS, 106, 306

\bibitem[\protect\citeauthoryear{Efstathiou et~al.}{1988}]{Efstathiou88}
Efstathiou, G., Frenk, C., White, S., \& Davis, M. 1988, MNRAS, 235, 715

\bibitem[\protect\citeauthoryear{Eisenstein \& Hut}{1998}]{HOP}
Eisenstein, D., \& Hut, P. 1998, ApJ, 498, 137

\bibitem[\protect\citeauthoryear{Evans \& Bridle}{2009}]{Evans09}
Evans, A., \& Bridle, S., 2009, ApJ, 695, 1446

\bibitem[\protect\citeauthoryear{{Faltenbacher} \& {White}}{2010}]{Faltenbacher09}
{Faltenbacher}, A., {White}, S. 2010, ApJ, 708, 469

\bibitem[\protect\citeauthoryear{Faltenbacher et~al.}{2002}]{Faltenbacher02}
Faltenbacher, A., Gottl\"{o}ber, S., Kerscher, M., \& M\"{u}ller, V. 2002, A\&A, 395, 1

\bibitem[\protect\citeauthoryear{Fischer et~al.}{2000}]{Fischer00}
Fischer, P., et~al. 2000, AJ, 120, 1198

\bibitem[\protect\citeauthoryear{Frenk et~al.}{1988}]{Frenk88}
Frenk, C., White, S., Davis, M., \& Efstathiou, G. 1988, ApJ, 327, 507

\bibitem[\protect\citeauthoryear{Frenk}{2002}]{Frenk02}
Frenk, C. 2002, Royal Society of London Philosophical Transactions Series A, 360, 1277

\bibitem[\protect\citeauthoryear{Gao \& White}{2007}]{Gao07}
Gao, L., \& White, S. 2007, MNRAS 377, L5

\bibitem[\protect\citeauthoryear{{Grief} {et~al.}}{2007}]{grief07}
Grief, T.~H., Johson, J.~L., Bromm, V., \& Klessen, R.~S. 2007, ApJ, 670, 1

\bibitem[\protect\citeauthoryear{Hahn et~al.}{2007}]{Hahn07}
Hahn O., Carollo, C., Porciani, C., \& Dekel, A. 2007, MNRAS, 381, 41

\bibitem[\protect\citeauthoryear{Harker et~al.}{2006}]{Harker06}
Harker, G., Cole, S., Helly, J., Frenk, C., \& Jenkins, A. 2006, MNRAS, 367, 1039

\bibitem[\protect\citeauthoryear{Hayashi et~al.}{2003}]{Hayashi03}
Hayashi, E., Navarro, J., Taylor, J., Stadel, J., \& Quinn, T., 2003, ApJ, 584, 541

\bibitem[\protect\citeauthoryear{Heitmann et~al.}{2006}]{Heitmann06}
Heitmann, K., Lukic, Z., Habib, S., \& Ricker, P. 2006, ApJ, 642, 85

\bibitem[\protect\citeauthoryear{{Hetznecker} \& {Burkert}}{2006}]{Hetz06}
{Hetznecker}, H., \& {Burkert}, A. 2006, 370, 1905

\bibitem[\protect\citeauthoryear{Hoekstra et~al.}{2003}]{Hoekstra03}
{Hoekstra}, H., Franx, M., Kuijken, K., Carlberg, R., \& Yee, H. 2003, MNRAS, 340, 609

\bibitem[\protect\citeauthoryear{{Hoyle}}{1949}]{hoyle49}
{Hoyle}, F. 1949, in Problems of Cosmical Aerodynamics. Central Air Documents, Office, Dayton, OH, p. 195

\bibitem[\protect\citeauthoryear{{Jang-Condell} \& {Hernquist}}{2001}]{jch01}
{Jang-Condell}, H., \& {Hernquist}, L. 2001, ApJ, 548, 68

\bibitem[\protect\citeauthoryear{Jing, Suto \& Mo}{2007}]{Jing07}
Jing, Y., Suto, Y., \& Mo, H. 2007, ApJ, 657, 664

\bibitem[\protect\citeauthoryear{{Johnson}, {Grief}, \& {Bromm}}{2007}]{Johnson07}
{Johnson}, J.~L., {Grief}, T.~H., \& {Bromm}, V. 2007, ApJ, 665, 85

\bibitem[\protect\citeauthoryear{Katz, Hernquist \& Weinberg}{1999}]{Katz99}
Katz, N., Hernquist, L., \& Weinberg, D. 1999, ApJ, 523, 463

\bibitem[\protect\citeauthoryear{Kauffmann et~al.}{1999}]{Kauffmann99}
Kauffmann, G., Colberg, J., Diaferio, A. \& White, S. 1999, MNRAS, 303, 188

\bibitem[\protect\citeauthoryear{Kravtsov, Gnedin, \& Klypin}{2004}]{Kravtsov04}
Kravtsov, A., Gnedin, O., \& Klypin, A., 2004, ApJ, 609, 482

\bibitem[\protect\citeauthoryear{Limousin et~al.}{2007}]{Limousin07}
Limousin, M., et~al. 2007, A\&A, 461, 881

\bibitem[\protect\citeauthoryear{{Macci{\'o}} {et~al.}}{2007}]{Maccio07}
{Macci{\'o}}, A., {Dutton}, A., {van den Bosch}, 
F., {Morre}, B., {Potter}, D., {Stadel}, J. 2007, MNRAS, 378, 55

\bibitem[\protect\citeauthoryear{{Macci{\'o}}, {Dutton}, \& {van den Bosch}}{2008}]{Maccio08}
{Macci{\'o}}, A., {Dutton}, A., {van den Bosch}, F. 2008, 391, 1940

\bibitem[\protect\citeauthoryear{{Maller}, {Dekel}, \& {Somerville}}{2002}]{maller02}
{Maller}, A., {Dekel}, A., \& {Somerville}, R. 2002, MNRAS, 329, 423

\bibitem[\protect\citeauthoryear{Mandelbaum et~al.}{2006}]{Mandelbaum06}
Mandelbaum, R., Hirata, C., Broderick, T., Seljak, U., \& Brinkman, J. 2006, MNRAS, 370, 1008

\bibitem[\protect\citeauthoryear{McKay et~al.}{2002}]{McKay02}
McKay, T., et~al. 2002, ApJ, 571, 85

\bibitem[\protect\citeauthoryear{{Mo} \& {White}}{2002}]{mo02}
{Mo}, H. \& {White}, S. 2002, MNRAS, 336, 112

\bibitem[\protect\citeauthoryear{Moore et~al.}{2006}]{Moore06}
Moore,B., Diemand, J., Madau, P., Zemp, M., \& Stadel, J. 2006 MNRAS, 368, 563

\bibitem[\protect\citeauthoryear{Navarro, Frenk, \& White}{1996}]{NFW96}
Navarro, J., Frenk, C., \& White, S. 1996, ApJ, 462, 563

\bibitem[\protect\citeauthoryear{Navarro, Frenk, \& White}{1997}]{NFW97}
Navarro, J., Frenk, C., \& White, S. 1997, ApJ, 490, 493

\bibitem[\protect\citeauthoryear{Neto et~al.}{2007}]{Neto07}
Neto, A. et~al. 2007, MNRAS, 381, 1450

\bibitem[\protect\citeauthoryear{Oesch et~al.}{2010}]{Oesch10}
Oesch, P. A. et~al. 2010, ApJ, 709, 16

\bibitem[\protect\citeauthoryear{Parker et~al.}{2007}]{Parker07}
Parker, L., Hoekstra, H., Hudson, M., van Waerbeke, L., \& Mellier Y. 2007, ApJ, 669, 21

\bibitem[\protect\citeauthoryear{{Peebles}}{1969}]{peebles69}
{Peebles}, P.~J.~E., 1969, ApJ , 155, 393 

\bibitem[\protect\citeauthoryear{Persic, Salucci, \& Stel}{1996}]{Persic96}
Persic, M., Salucci, P., \& Stel, F. 1996, MNRAS, 281, 27

\bibitem[\protect\citeauthoryear{{Porciani}, {Dekel} \& {Hoffman}}{2002}]{porci02}
{Porciani}, C., {Dekel}, A., \& {Hoffman}, Y. 2002, MNRAS, 332, 325

\bibitem[\protect\citeauthoryear{Press \& Schechter}{1974}]{PS74}
Press, W., \& Schechter, P. 1974, ApJ, 187, 425

\bibitem[\protect\citeauthoryear{{Prunet} {et~al.}}{2008}]{mpgrafic}
{Prunet}, S., {Pichon}, C., {Aubert}, D., {Pogosyan}, D., {Teyssier}, R., \& {Gottloeber}, S. 2008, ApJS, 178, 179

\bibitem[\protect\citeauthoryear{Quadri, M\"{o}ller, \& Natarajan}{2003}]{Quadri03}
Quadri, R., M\"{o}ller, O., \& Natarajan, P. 2003, ApJ, 597, 659.

\bibitem[\protect\citeauthoryear{Quadri et~al.}{2008}]{Quadri08}
Quadri, R., Williams, R., Lee, K.-S., Franx, M., van Dokkum, P., \& Brammer, G. 2008, ApJ, 685, 1

\bibitem[\protect\citeauthoryear{{Reed} {et~al.}}{2009}]{reed08}
{Reed}, D., {Bower}, R., {Frenk}, C., {Jenkins}, A., \& {Theuns}, T. 2009, MNRAS, 394, 624

\bibitem[\protect\citeauthoryear{{Reed} {et~al.}}{2007}]{reed07}
{Reed}, D., {Bower}, R., {Frenk}, C., {Jenkins}, A., \& {Theuns}, T. 2007, MNRAS, 374, 2

\bibitem[\protect\citeauthoryear{Reed et~al.}{2005}]{Reed05}
Reed, D., Governato, F., Quinn, T., Gardner, J., Stadel, J., \& Lake, G. 2005, MNRAS 359, 1537

\bibitem[\protect\citeauthoryear{{Reed} {et~al.}}{2003}]{reed03}
{Reed}, D., et~al. 2003, MNRAS, 346, 565

\bibitem[\protect\citeauthoryear{Rubin et~al.}{1985}]{Rubin85}
Rubin, V., Burnstein, D., Ford, W., \& Thonnard, N. 1985, ApJ, 289, 81

\bibitem[\protect\citeauthoryear{Salucci et~al.}{2007}]{Salucci07}
Salucci, P., Lapi, A., Tonini, C., Gentile, G., Yegorova, I, \& Klein, U. 2007 MNRAS, 378, 41

\bibitem[\protect\citeauthoryear{Shaw et~al.}{2006}]{Shaw06}
Shaw, L., Weller, J., Ostriker, J. \& Bode, P. 2006, ApJ, 646, 815

\bibitem[\protect\citeauthoryear{{Sheth} \& {Tormen}}{1999}]{st99}
{Sheth}, R., \& {Tormen}, G. 1999, MNRAS, 308, 119

\bibitem[\protect\citeauthoryear{{Spergel} {et~al.}}{2007}]{wmap3}
{Spergel}, D.~N., et~al. 2007, ApJS, 170, 377

\bibitem[\protect\citeauthoryear{{Springel}}{2005}]{Gadget05}
{Springel}, V. 2005, MNRAS, 364, 1105

\bibitem[\protect\citeauthoryear{Springel}{2005}]{Springel05}
Springel, V. et~al. 2005, Nature, 435, 629

\bibitem[\protect\citeauthoryear{{Sugerman}, {Summers}, \& {Kamionkowski}}{2000}]{sugerman00}
{Sugerman}, B., {Summers}, F., \& {Kamionkowski}, M. 2000, MNRAS, 311, 762

\bibitem[\protect\citeauthoryear{{Stadel}}{2001}]{skid}
{Stadel}, J.~G. 2001, Ph.D. thesis, Univ. Washington

\bibitem[\protect\citeauthoryear{{Steinmetz} \& {Bartelmann}}{1995}]{steinmetz95}
{Steinmetz}, M., \& {Bartelmann}, M. 1995, MNRAS, 272, 570

\bibitem[\protect\citeauthoryear{Trenti et~al.}{2010}]{Trenti10}
Trenti, M., Smith, B., Hallman, E., Skillman, S., \& Shull, J. 2010, ApJ, 711, 1198

\bibitem[\protect\citeauthoryear{Trimble}{1987}]{Trimble87}
Trimble, V. 1987 ARA\&A 25, 425

\bibitem[\protect\citeauthoryear{{Vitvitska} {et~al.}}{2002}]{vitvit02}
{Vitvitska}, M., {Klypin}, A.~A., {Kravtsov}, A.~V., {Wechsler}, R.~H.,{Primack}, J.~R., \& {Bullock}, J.~S. 2002, ApJ, 581, 799

\bibitem[\protect\citeauthoryear{{Warren} {et~al.}}{1992}]{warren92}
{Warren}, M.~S., {Quinn}, P.~J., {Salmon}, J.~K.,\& {Zurek}, W.~H. 1992, ApJ, 399, 405

\bibitem[\protect\citeauthoryear{Warren et~al.}{2006}]{Warren06}
Warren, M., Abazaijian, K., Hold, D., \& Teodoro, L. 2006, ApJ, 646, 881

\bibitem[\protect\citeauthoryear{Wechsler et~al.}{2006}]{Wechsler06}
Wechsler, R., Zentner, A., Bullock, J., Kravtsov, A., \& Allgood, B. 2006, ApJ, 652, 71

\bibitem[\protect\citeauthoryear{Wetzel et~al.}{2007}]{Wetzel07}
Wetzel, A., Cohn, J., White, M., Holz, D., \& Warren, M. 2007, ApJ, 656, 139

\bibitem[\protect\citeauthoryear{{Whalen} {et~al.}}{2008a}]{Whalen08A}
Whalen, D., van Veelen, B., O'Shea, B., \& Norman, M., 2008, ApJ, 682, 49

\bibitem[\protect\citeauthoryear{{Whalen} {et~al.}}{2008b}]{Whalen08B}
Whalen, D., O'Shea, B., Smidt, J., \& Norman, M., 2008, ApJ, 679, 925

\bibitem[\protect\citeauthoryear{{White}}{1984}]{white84}
{White}, S.~D.~M., 1984, ApJ, 286, 38

\end{thebibliography}
\end{document}